\documentclass[
    ,final            
  ]
  {aipproc}

\layoutstyle{8x11single}

\newcommand{\q}[2]{\ensuremath{#1\ \mathrm{#2}}} 
\newcommand{\vol}[1]{\textbf{#1}} 
\newcommand{\kick}{\ensuremath{\theta}}
\newcommand{\betaCS}{\ensuremath{\beta}} 
\newcommand{\Brho}{\ensuremath{(B\rho)}} 
\newcommand{\Aring}{\ensuremath{A_\mathrm{ring}}} 
\newcommand{\Amax}{\ensuremath{A_\mathrm{max}}} 

\begin{document}

\title[Applications of electron lenses]{Applications of electron
  lenses: scraping of high-power beams, beam-beam compensation, and
  nonlinear optics}


\author{Giulio Stancari}{ address={Fermi National Accelerator
    Laboratory, Batavia, Illinois 60510, U.S.A.\thanks{Fermi National
      Accelerator Laboratory (Fermilab) is operated by Fermi Research
      Alliance, LLC under Contract No.~DE-AC02-07CH11359 with the
      United States Department of Energy.  This work was partially
      supported by the US DOE LHC Accelerator Research Program (LARP)
      and by the European FP7 HiLumi LHC Design Study, Grant Agreement
      284404. Report number: FERMILAB-CONF-14-314-APC.}}
}

\begin{abstract}
  Electron lenses are pulsed, magnetically confined electron beams
  whose current-density profile is shaped to obtain the desired effect
  on the circulating beam. Electron lenses were used in the Fermilab
  Tevatron collider for bunch-by-bunch compensation of long-range
  beam-beam tune shifts, for removal of uncaptured particles in the
  abort gap, for preliminary experiments on head-on beam-beam
  compensation, and for the demonstration of halo scraping with hollow
  electron beams. Electron lenses for beam-beam compensation are being
  commissioned in the Relativistic Heavy Ion Collider (RHIC) at
  Brookhaven National Laboratory (BNL). Hollow electron beam
  collimation and halo control were studied as an option to complement
  the collimation system for the upgrades of the Large Hadron Collider
  (LHC) at CERN; a conceptual design was recently completed. Because
  of their electric charge and the absence of materials close to the
  proton beam, electron lenses may also provide an alternative to
  wires for long-range beam-beam compensation in LHC luminosity
  upgrade scenarios with small crossing angles. At Fermilab, we are
  planning to install an electron lens in the Integrable Optics Test
  Accelerator (IOTA, a 40-m ring for 150-MeV electrons) as one of the
  proof-of-principle implementations of nonlinear integrable optics to
  achieve large tune spreads and more stable beams without loss of
  dynamic aperture.
\end{abstract}

\date{\today}

\maketitle


\section{Introduction}

Electron lenses are pulsed, magnetically confined, low-energy electron
beams whose electromagnetic fields are used for active manipulation of
the circulating beam in high-energy
accelerators~\cite{Shiltsev:PRSTAB:2008, Shiltsev:Handbook:2013}. The
first main feature of an electron lens is the possibility to control
the current-density profile of the electron beam (flat, Gaussian,
hollow, etc.) by shaping the cathode and the extraction
electrodes. Another feature is pulsed operation, enabled by the
availability of high-voltage modulators with fast rise times. The
electron beam can therefore be synchronized with subsets of bunches,
with different intensities for each subset. The main advantage of the
use of electron lenses for high-power accelerators is the absence of
metal close to the beam, therefore avoiding material damage and
impedance.
Electron lenses were developed for beam-beam compensation in
colliders~\cite{Shiltsev:elens-bbcomp}, enabling the first observation
of long-range beam-beam compensation effects by tune shifting
individual bunches~\cite{Shiltsev:PRL:2007}. They were used for many
years during regular Tevatron collider operations for cleaning
uncaptured particles from the abort gap~\cite{Zhang:PRSTAB:2008}. One
of the two Tevatron electron lenses was used for experiments on
head-on beam-beam compensation in 2009~\cite{Stancari:BB:2013}, and
for exploring hollow electron beam collimation in
2010--2011~\cite{Stancari:PRL:2011, Stancari:APSDPF:2011}. Electron
lenses for beam-beam compensation were built for RHIC at BNL and are
being commissioned~\cite{Fischer:IPAC:2013, Fischer:IPAC:2014}.
Current areas of research on electron lenses include the generation of
nonlinear integrable lattices in IOTA at
Fermilab~\cite{Nagaitsev:IPAC:2012, Valishev:IPAC:2012} and
applications for the LHC upgrades: as halo monitors and scrapers, as
charged current-carrying `wires' for long-range beam-beam
compensation, and as tune-spread generators for Landau damping of
instabilities before collisions.

\section{Nonlinear integrable optics with electron lenses}

\begin{figure}[b!]
\includegraphics[width=\textwidth]{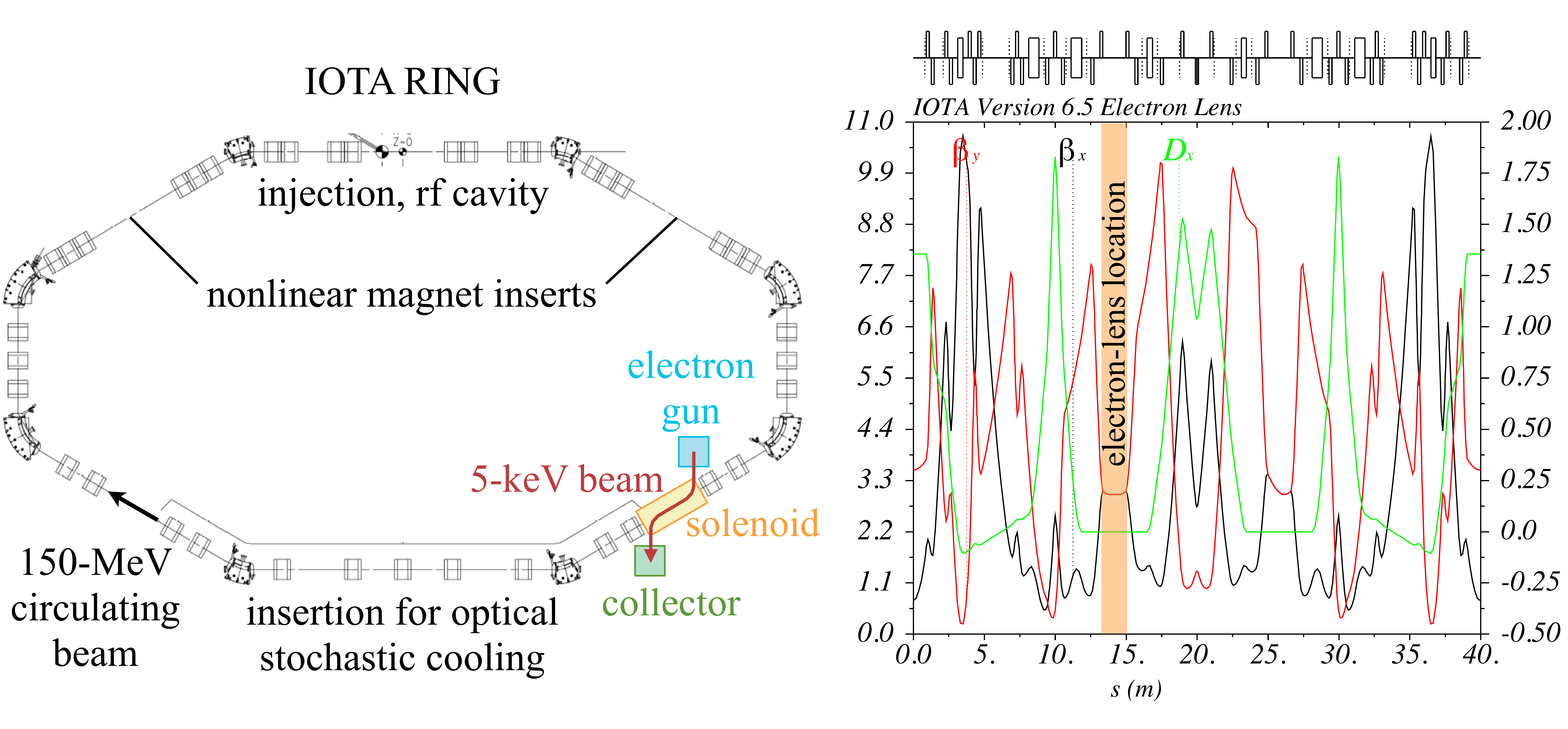}
\source{G.~Kafka (IIT/Fermilab) and A.~Valishev (Fermilab)}
\caption{Left: schematic layout of the IOTA ring; the electron lens
  (gun, solenoid, and collector) is shown in the lower right
  section. Right: calculated IOTA lattice for electron-lens operation,
  showing the horizontal and vertical amplitude functions $\beta_x$
  and $\beta_y$ (left axis, in meters) and the horizontal
  dispersion~$D_x$ (right axis, in meters) as a function of the
  longitudinal coordinate~$s$ around the ring.}
\label{fig:IOTA}
\end{figure}

The Integrable Optics Test Accelerator (IOTA) is a small storage ring
(40~m circumference) being built at Fermilab as part of the
accelerator and beam physics research program. Some of the main goals
of the project are the practical implementation of nonlinear
integrable lattices in a real machine, the study of space-charge
compensation in rings, and a proof-of-principle demonstration of
optical stochastic cooling~\cite{Nagaitsev:IPAC:2012,
  Valishev:IPAC:2012}.

The concept of nonlinear integrable optics applied to accelerators
involves a small number of special nonlinear focusing elements added
to the lattice of a conventional machine in order to generate large
tune spread while preserving dynamic
aperture~\cite{Danilov:PRSTAB:2010}. The concept may have a profound
impact in the design of high-intensity machines by providing improved
stability to perturbations and mitigation of collective instabilities
through Landau damping.

The effect of nonlinear lattices on single-particle dynamics will be
investigated during the first stage of IOTA operations using
low-intensity pencil beams of electrons at 150~MeV:
\q{10^9}{particles/bunch}, \q{0.1}{\mu m} transverse rms geometrical
equilibrium emittance, 2~cm rms bunch length, and $1.4\times 10^{-4}$
relative momentum spread. The beam is generated by the photoinjector
currently being commissioned at the Fermilab Advanced Superconducting
Test Accelerator (ASTA) facility.  The goal of the project is to
demonstrate a nonlinear tune spread of about 0.25 without loss of
dynamic aperture in a real accelerator.

It was recently shown that one way to generate a nonlinear integrable
lattice is with specially segmented quadrupole
magnets~\cite{Danilov:PRSTAB:2010}. There are also 2~concepts based on
electron lenses: (a)~axially symmetric thin-lens kicks with a
particular amplitude dependence~\cite{McMillan:1967, McMillan:1971,
  Danilov:PAC:1997}; and (b)~axially symmetric thick-lens kicks in a
solenoid~\cite{Nagaitsev:private}. These concepts use the
electromagnetic field generated by the electron beam distribution to
provide the desired nonlinear transverse kicks to the circulating
beam.

The integrability of axially symmetric thin-lens kicks was studied in
1~dimension by McMillan~\cite{McMillan:1967, McMillan:1971}. It was
then extended to 2~dimensions~\cite{Danilov:PAC:1997} and
experimentally tested with colliding beams~\cite{Shwartz:BB:2013}. Let
us analyze the main quantities involved in the electron-lens case. The
beam in the electron lens (Figure~\ref{fig:IOTA}, left) has velocity
$v_e = \beta_e c$. The length~$L$ of the electron lens is assumed to
be small in comparison with the local lattice amplitude
function~\betaCS. Let $j(r)$ be a specific radial dependence of the
current density of the electron-lens beam (the `McMillan case'), with
$j_0$ its value on axis and $a$ its effective radius:
$j(r) = j_0 a^4 / (r^2 + a^2)^2$.
The total current is
$I_e = 2 \pi \int_{0}^{\infty} j \cdot r \, dr = j_0 \pi a^2$.
While traversing the electron lens, the circulating beam, with
magnetic rigidity~\Brho\ and velocity $v_z = \beta_z c$, experiences
the following transverse angular kick:
\begin{equation}
\kick(r) = 2\pi \frac{j_0 L (1 \pm \beta_e \beta_z)}
                     {\Brho \beta_e \beta_z c^2}
  \frac{a^2 r}{r^2 + a^2}
  \left( \frac{1}{4\pi\epsilon_0} \right) .
\label{eq:McMillan-kicks}
\end{equation}
The `$+$' sign applies when the beams are counterpropagating and the
electric and magnetic forces act in the same direction. For such a
radial dependence of the kick, there are 2~noncommuting invariants of
motion in the 4-dimensional transverse phase space. Neglecting
longitudinal effects, all particle trajectories are regular and
bounded.

The concept of axially symmetric thick-lens kicks relies on a solenoid
with axial field~$B_z = 2 \Brho / \betaCS$ to provide focusing for the
circulating beam and constant amplitude lattice functions $\betaCS
\equiv \beta_x = \beta_y$. The same solenoid magnetically confines the
low-energy beam in the electron lens. In this case, any axially
symmetric electron-lens current distribution $j(r)$ generates
2~conserved quantities (the Hamiltonian and the longitudinal component
of the angular momentum), as long as the betatron phase advance in the
rest of the ring is an integer multiple of $\pi$.

The achievable nonlinear tune spread~$\Delta\nu$ (i.e., the tune difference
between small and large amplitude particles) is proportional to the
electron-lens current density on axis:
\begin{equation}
\Delta\nu = \frac{\betaCS j_0 L (1\pm \beta_e \beta_z)}
                 {2 \Brho \beta_e \beta_z c^2}
  \left(\frac{1}{4\pi\epsilon_0}\right).
\label{eq:dnu}
\end{equation}

For demonstrating the nonlinear integrable optics concept with
electron lenses in a real machine, there are several design
considerations to take into account.

The size of the electron beam should be compatible with the achievable
resolution of the apparatus. Amplitude detuning and dynamic aperture
of the ring will be measured by observing the turn-by-turn position
and intensity of a circulating pencil beam with an equilibrium
emittance $\epsilon = \q{0.1}{\mu m}$ (rms, geometrical) and size
$\sigma_e = \sqrt{\betaCS \epsilon}$ at the electron lens. This size
should be larger than the expected resolution of the beam position
monitors, $\sigma_\mathrm{BPM} \leq \q{0.1}{mm}$. In the current IOTA
lattice design (Figure~\ref{fig:IOTA}, right), $\betaCS = \q{3}{m}$
and $\sigma_e = \sqrt{\betaCS \epsilon} = \q{0.55}{mm}$, which
satisfies this requirement. Moreover, it follows that the required
axial field is $B_z = 2 \Brho / \betaCS = \q{0.33}{T}$.

The aperture of the ring $\Aring = \q{24}{mm}$ must be sufficient to
contain a wide range of betatron amplitudes and detunigs. Aperture and
magnet field quality suggest a maximum tolerable orbit excursion of
about $\Amax = \Aring/2 = \q{12}{mm}$. Particles at small amplitudes
will exibit the maximum detuning $\Delta\nu$. The maximum excursion
\Amax\ must be sufficient to accomodate particles with large
amplitudes and small detunings compared to $\Delta\nu$. For the
McMillan kick distribution of Eq.~\eqref{eq:McMillan-kicks}, for
instance, this can be achieved by requiring $a \leq \Amax/6 =
\q{2}{mm}$. For a typical electron lens with resistive solenoids, with
$B_z = \q{0.33}{T}$ in the main solenoid, one can operate at $B_g =
\q{0.1}{T}$ in the gun solenoid. Because of magnetic compression, this
translates into a current-density distribution with $a_g = a
\sqrt{B_z/B_g} = \q{3.6}{mm}$ at the cathode. This parameter serves as
an input to the design of the electron-gun assembly.

The achievable tune spread should be large enough to clearly
demonstrate the effect. As a goal for the IOTA project, it was decided
to set $\Delta\nu \geq 0.25$. Through Eq.~\eqref{eq:dnu}, this
requirement imposes a constraint on the current density in the
electron lens.
%
%
For instance, with typical electron-lens parameters, $L = \q{1}{m}$,
$\beta_e = 0.14$ (5~keV kinetic energy) and copropagating
beams\footnote{At the design stage, we consider the copropagating case
  because it is more conservative and because it may serve in the
  future as an electron cooler for protons, which will be injected in
  IOTA in the same direction as electrons.}, one obtains $j_0 =
\q{14}{A/cm^2}$ and, for the McMillan distribution, a total current
$I_e = j_0 \pi a^2 = \q{1.7}{A}$.

The design parameters are within the current state of the art. It may
be challenging to transport these currents through a resistive
electron lens while preserving the desired quality of the
current-density profile. The effects of imperfections and of
longitudinal fields must be investigated with numerical simulations
and with experimental studies in the Fermilab electron-lens test
stand. The relatively large instantaneous beam power to be dissipated
in the collector (8.5~kW) may require provisions for pulsed operation
of the electron lens compatible with the time structure of the IOTA
circulating beam. In general, the project benefits from the many years
of experience in the construction and operation of electron lenses at
Fermilab, and it can rely on several components that are already
available at the laboratory, such as electron gun assemblies,
resistive solenoids, collectors, and power supplies.

\section{Halo control with hollow electron beams}


Hollow electron beam collimation is the most mature among the
electron-lens applications discussed here. A recent summary can be
found in Ref.~\cite{Stancari:IPAC:2014}. The technique was tested in
the Fermilab Tevatron collider and it is now being proposed as an
option to complement the LHC collimation system. It is based upon
electron beams with a hollow current-density profile aligned with the
circulating beam~\cite{Shiltsev:HEBC, Stancari:PRL:2011,
  Stancari:APSDPF:2011}.  If the electron distribution is axially
symmetric, the proton beam core is unperturbed, whereas the halo
experiences smooth and tunable nonlinear transverse kicks. The size,
position, intensity, and time structure of the electron beam can be
controlled over a wide range of parameters.


For the LHC and its luminosity upgrades (HL-LHC), beam halo
measurement and control are critical, and this technique may provide
unique capabilites. LHC and HL-LHC represent huge leaps in stored beam
energy.
Beam halo monitoring and control are one of the major risk factors for
LHC and for safe operation with crab cavities in HL-LHC. There is a need to
measure and monitor the beam halo, and to remove it at controllable
rates. Hollow electron lenses are the most established and flexible
tool for this purpose.

A plan for electron lenses in the LHC was
developed
and a conceptual design
was recently completed~\cite{Stancari:PAC:2013,
  Stancari:CDR:2014}. The expected performance is based upon
experimental measurements and numerical
simulations~\cite{Valishev:TM:2014}.
A wide range of halo removal
rates is possible, from seconds to hours, using the electron lens in
continuous mode (same electron current every turn for a given bunch)
or in stochastic mode (by adding random noise turn by turn to the
modulator voltage).

Alternative schemes for halo control will also
be investigated, as they may be cheaper than electron lenses, or they
may become available sooner. These alternatives include excitations
with transverse dampers, tune modulation with warm
quadrupoles~\cite{Bruening:tunemod}, and wire compensators.
Noninvasive halo diagnostic methods, such as synchrotron-light
monitoring with wide dynamic range, are being pursued with high
priority. The electron lenses themselves, if they are installed, may
provide a new sensitive way to measure halo populations with
backscattered electron detectors, as is being demonstrated at
RHIC~\cite{Fischer:IPAC:2014, Thieberger:BIW:2012}.

\section{Electron `wires' as long-range compensators}

Electron lenses may play an important role in HL-LHC luminosity
schemes with flat beams and smaller crossing angles, where no crab
cavities are necessary, but for which long-range beam-beam
compensation is critical.
Conventional wire compensators will be tested after the current LHC
shutdown. They are technically challenging and they present a risk for
collimation and machine protection, because they involve water-cooled
copper cables carrying 378~A at about 10 standard deviations of the
proton beam size from the beam axis. Electron lenses are considered as
a safer, less demanding alternative to wire compensators, with the
added benefit of pulsing~\cite{Valishev:TM:2013}. About 21~A over a
distance of 3~m would be required for HL-LHC, with any transverse
shape.
Physics and integration studies were initiated to calculate the
expected performance and its sensitivity to location. Energy
deposition in the superconducting solenoid and radiation to the
high-voltage modulator must also be addressed.

\section{Conclusions}

Electron lenses are unique devices for active beam manipulation in
accelerators, with a wide range of applications. Halo scraping with
hollow electron beams was demonstrated at the Fermilab Tevatron
collider and, because halo measurement and control are critical for
LHC and its upgrades, a conceptual design of hollow electron beam
scraper for the LHC was recently completed. Electron lenses in the LHC
are also a candidate for long-range beam-beam compensation. Magnetized
low-energy electron beams are also relevant for the Fermilab beam
physics program in the near future, as one of the most promising ways
to practically implement nonlinear integrable lattices in the IOTA
ring.


\begin{theacknowledgments}
  S.~Nagaitsev and A.~Valishev (Fermilab) are leading the research in
  nonlinear integrable optics at Fermilab. They developed most of the
  theoretical concepts, designs, and practical solutions for the IOTA
  ring. G.~Kafka (IIT/Fermilab) designed lattices for various IOTA
  operation modes as part of his Ph.D.\ thesis work.
  The LHC Collimation Team led the
  conceptual and technical design of electron lenses for the LHC:
  in particular, R.~Bruce, S.~Redaelli, A.~Rossi, and B.~Salvachua
  Ferrando.
  The contributions, ideas, support, and practical advice of the
  following people were greatly appreciated: O.~Aberle, A.~Bertarelli,
  F.~Bertinelli, O.~Br\"uning, G.~Bregliozzi, P.~Chiggiato,
  S.~Claudet, R.~Jones, Y.~Muttoni, D.~Perini, L.~Rossi, B.~Salvant,
  H.~Schmickler, R.~Steinhagen, G.~Tranquille, G.~Valentino (CERN),
  V.~Moens (EPFL), G.~Annala, G.~Apollinari, M.~Chung, T.~Johnson,
  I.~Morozov, E.~Prebys, V.~Previtali, G.~Saewert, V.~Shiltsev,
  D.~Still, L.~Vorobiev (Fermilab), R.~Assmann (DESY), M.~Blaskiewicz,
  W.~Fischer, X.~Gu (BNL), D.~Grote (LLNL), H.~J.~Lee (Pusan National
  U., Korea), S.~Li (Stanford U.), A.~Kabantsev (UC San Diego),
  T.~Markiewicz (SLAC), and D.~Shatilov (BINP).
\end{theacknowledgments}



\bibliographystyle{aipproc}   



\end{document}